# Energy-Aware Virtual Network Embedding Approach for Distributed Cloud

Amal S. Alzahrani
Collage of Computer Engineering and Science
Prince Sattam Bin Abdulaziz university
Riyadh, Kingdom of Saudi Arabia

Ashraf A. Shahin[1,2]
[1]College of Computer and Information Sciences,
Al Imam Mohammad Ibn Saud Islamic University (IMSIU)
Riyadh, Kingdom of Saudi Arabia
[2]Department of Computer, Institute of Statistical Studies & Research, Cairo University, Cairo, Egypt

*Abstract*—Network virtualization has caught the attention of many researchers in recent years. It facilitates the process of creating several virtual networks over a single physical network. Despite this advantage, however, network virtualization suffers from the problem of mapping virtual links and nodes to physical network in most efficient way. This problem is called virtual network embedding ("VNE"). Many researches have been proposed in an attempt to solve this problem, which have many optimization aspects, such as improving embedding strategies in a way that preserves energy, reducing embedding cost and increasing embedding revenue. Moreover, some researchers have extended their algorithms to be more compatible with the distributed clouds instead of a single infrastructure provider ("ISP"). This paper proposes energy aware particle swarm optimization algorithm for distributed clouds. This algorithm aims to partition each virtual network request ("VNR") to sub-graphs, using the Heavy Clique Matching technique ("HCM") to generate a coarsened graph. Each coarsened node in the coarsened graph is assigned to a suitable data center ("DC"). Inside each DC, a modified particle swarm optimization algorithm is initiated to find the near optimal solution for the VNE problem. The proposed algorithm was tested and evaluated against existing algorithms using extensive simulations, which shows that the proposed algorithm outperforms other algorithms.

*Keywords*—*Distributed virtual network embedding; energy consumption; particle swarm optimization; network virtualization; virtual network embedding; virtual network request; virtual network partitioning*

## I. INTRODUCTION

Cloud computing is a computational paradigm that deliver on demand, pay as you use services. These services include Software as a Service (SaaS) by allowing users to use application over internet, Platform as a Service (PaaS) as operating system, databases and web servers and Infrastructure as a Service (IaaS), such servers and software. In order to deliver these services, each cloud should encounter many resources, such servers to fulfill users' demands, where each service will be dedicated to single user at time, thus increasing service cost and power consumption. On the other hand, failure in one server will have consequences on overall services provided.

One of the most important feature of cloud computing is virtualization. It is a method of logically partition physical resources in a way that one physical resource can accommodate multiple users' demands at same time. As a result, sharing resources will help to reduce cost and energy consumption along with increasing resources utilization [1].

As part of virtualization, network virtualization caught attention of many researchers during the past few years. It facilitates the process of creating several virtual networks over a single physical network called Substrate Network "SN". It provides resources sharing requirement over cloud computing infrastructure. Network Virtualization plays an important role as link between virtual and physical infrastructure. Therefore, the process of virtual resources allocation over the corresponding physical ones became a critical issue. This problem called Virtual Network Embedding "VNE". It defined as the problem of mapping virtual nodes and links to physical nodes and paths [2].

The VNE problem can be divided into two stages: node mapping and link mapping. Node mapping is where each virtual node in the VN request should map to the corresponding one in the SN. The virtual node resource requirements never exceed what the physical node can offer. After this stage, link mapping takes place. Link mapping is more complicated than node mapping, the reason being that the virtual link should map to the physical path while keeping within the bandwidth ("BW") constraints.

However, mapping virtual nodes and links in separate stages increases the embedding cost. This is because mapping two neighboring virtual nodes far away from each other increases the length of the substrate path.

The main concern regarding VNE problem is how to map virtual resource to physical one while maintaining lower embedding cost. Beside the importance of reducing the embedding cost, the energy consumption during mapping stage should be taken in consideration. Energy conservation means to utilize the SN resources in way that saves overall power. This can be done through switching off the underutilizing resources or migrating VN requests [3]-[6]. On the other hand, embedding VN requests over multiple domains had its fair share of attention [7], [8].

Embedding request can be solved in single ISP. In this situation it called intra-domain. In contrast, inter-domain mapping as in distributed clouds computing is considered a substantial topic in most of the recent researches conducted in the area of computer technology. It derives its importance from its ability to handle large amounts of requests due to the size of the resources offered. It is composed of multiple Datacenters ("DCs") distributed geographically [9].

Because of the distributed nature, important questions must be asked. Where is the best to place the VN requests? Which DC is the best to fulfill the virtual network's demands? How can virtual requests be placed so that cost and energy consumption constraints are maintained?

From these points, this paper proposed Energy Aware Virtual Network embedding based on particle swarm optimization algorithm. It intended to provide an optimization approach that aims to minimizing both embedding cost along with energy consumption in distributed clouds.

The proposed optimization approach is based on adopting Particle swarm optimization "PSO" algorithm in VNE problem. PSO is a population based algorithm that starts with candidate solutions in one iteration, and improve them in the next one [10]. The performance of the proposed algorithm was tested against existing algorithms. It presents noticeable improvements regarding energy consumption, revenue, acceptance ratio, VNE time, achieved and rejected resources comparing to some of the existing algorithms.

The rest of this paper is organized as follows. Sections 2 and 3 give a short overview of research background and related work. Section 4 presents the VN embedding model and problem formulation. Section 5 describes the proposed algorithm. Section 6 evaluates the proposed VN embedding algorithm. Finally, research conclusion is in Section 7.

## II. BACKGROUND

This paper is based on two technologies that form ground of the presented research, Network Virtualization and Particle Swarm Optimization.

### A. Network Virtulization

Cloud computing can be defined as computing that delivers and permits access to shared resources over the Internet. Along with cloud computing, network virtualization considered a promising solution that accommodates the rapid growth of the Internet. It aims to support the creation of multiple virtual networks ("VNs") in the same shared physical network (Fig. 1).

A virtual network is composed of virtual nodes that are connected via virtual links. Network virtualization as an area is composed of two main players: Infrastructure providers (INPs), which are responsible for managing the physical infrastructure, and service providers (SPs), which are responsible for creating and maintaining VN requests [11]. However, despite this major flexibility that network virtualization brings, many questions have arisen regarding the problem of mapping the VN to the corresponding SN. This is called virtual network embedding and is defined as the problem of mapping virtual nodes and links to physical nodes and the corresponding path, respectively. It is considered an NP-hard problem [11].

Another problem that seems to have gained a lot of attention recently is energy consumption when mapping the VN. Many optimization solutions have been proposed over the last few years. These research projects aimed to solve the VNE problem with minimum embedding cost and better energy conserving strategies. Some of these proposed solutions followed the "Exact Solution" technique, where an optimal solution can be found through integer linear programming and mixed integer liner programming. However, such a technique is not Practicable in a large problem space. Therefore, heuristic and metaheuristic algorithms are presented as substitutes for traditional algorithms. These include genetic algorithms, simulated annealing, evolutionary programming and particle swarm optimization [2].

### B. Particle Swarm Optimization

This paper focuses in modifying PSO algorithm to solve VNE problem to reach the goal of reducing embedding cost and energy consumption. This algorithm was inspired by birds' flock movements. Considering this algorithm as metaheuristic algorithm make it more applicable in large search space. PSO is optimization algorithm aims to find the near optimal solution among other candidate ones [10], [11]. In PSO, each candidate solution considered as particle, where each particle has its own position "x" and velocity "v". The optimization accomplished by moving each particle's position toward the optimal solution. This can be done by changing the particle's position and velocity according the following equation [18]:

$$v_i^d = wv_i^d + c_1 r_1^d (pBest_i^d - x_i^d) + c_2 r_2^d (gBest^d - x_i^d) \quad (1)$$

$$x_i^d = x_i^d + v_i^d \quad (2)$$

Where, $pBest_i^d$ is the particle's best position and $gBest^d$ is the best position in the whole swarm. Both r1 and r2 are random values and c1, c2 are cognition and social weight and w is inertia weight.

## III. RELATED WORKS

As result to the importance of Network Virtualization, many researches were conducted to solve VNE problem [1] [12], [13]. In [12] Chevalier, C. and Safro, I. aimed to find near optimal solution for VNs embedding problems. First, they proposed a unified enhanced particle swarm optimization-based "VNE-UEPSO". The proposed algorithm focused on two stages: finding initial solutions "particles", and applying modified particle swarm optimization algorithm to find VNs embedding solution. Also, [6] presented embedding algorithm based on discrete particle swarm optimization. They modeled the VNE problem as binary optimization problem and proved by experiments that their algorithm consumes less computation time. In [14] author proposed a distributed algorithm for virtual networks mapping to substrate network resources. In addition, they proposed a VM mapping protocol to maintain communication and messages exchange between substrate nodes. The proposed algorithm has an impact on reducing the time delay in accepting multiple VN requests in same time. In [1], Kumar, R., and Charu, S. introduced a distributed protocol and VN embedding framework that ensure of mapping VN request across heterogeneous INPs. Additionally, they proposed "COST" addressing scheme for encoding and representing the geolocation information in each INP nodes. Another proposed protocol was Location Aware Protocol "LAP". It worked as assistance protocol that helped Inps in making informed decisions regarding which INP should forward the VN request. In [15] authors aimed to provide solutions regarding the problems associated to the energy consumption in VN embedding. Their focus was on determine when and which virtual machine should be migrated, and to witch host. In [16] authors presented the algorithm to minimize the power consumption by switching off as many substrate nodes as possible without affecting the overall network performance. Additionally, they proposed Mixed Integer Program to solve VN embedding problem which considered as NP-complete. As in [17] the proposed work focused on two phases. Firstly, VM placement, that is achieved by Modified Best Fit Decreasing "MBFD". This algorithm aimed to place VM with higher CPU utilization to a host that causes least power consumption. Second, VM selection which can be described as the process of choosing the VM that should be migrated to another host. From different point of view, [2] aimed to provide a solution for virtual network embedding in networked clouds. Their main contribution was mapping unbound virtual network request to different cloud sites. Their proposed solution consists of the following steps: 1) construct inter-domain physical graph; 2) construct cloud graph from the physical graph, where each node represents cloud site; 3) partitioned the virtual topology; and 4) find the solution with minimum embedding cost. In [3] authors proposed discovery framework for virtual resource organization across multiple infrastructure providers. The framework composed of two main parts: First, Management Nodes, "MN" responsible of maintaining and classifying the local virtual resource in each ISP to conceptual clusters. Each of these clusters named as Micro Cluster "MiC". Second, the Cluster Index Servers, "CIS" is responsible of organizing each MiC with the same root to represent Macro Clusters, "MaC". The proposed research's main contribution was to find discovery framework that facilitates the organization of virtual resource across different Inp. In [18] authors proposed research aimed to solve both nodes and links mapping problem in intra-domain network. Researchers proposed algorithm worked in sorting the nodes according to their resource requirements. In contrast to previous works, they included link mapping cost in the sorting process. In [4], author proposed algorithms for assigning substrate network resources to virtual network components. They focused on node stress ration in the assignment process, where node with highest stress ratio is selected as center node. In [5], author presented algorithm that depend on additional factor while calculating the stress used for mapping VN in to heterogeneous SN. These additional parameters include: link bandwidth, CPU load and frequency, candidate nodes for each Virtual node and free RAM. In [7], author proposed two mapping algorithms: Integer Liner Programming and Heuristic algorithm. Their main goals were minimizing CPU, RAM load and Link load. On the other hand, the proposed heuristic algorithm was similar to the one discussed in [5] addition to some modifications. It included both CPU and RAM consumption in node stress calculations. Tuning value was added to reduce link-path cost and update link stress after each VN mapping.

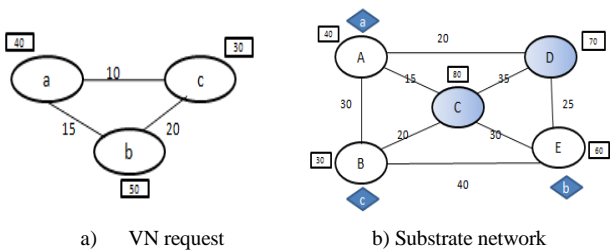

a) VN request  b) Substrate network

Fig. 1. Example of VN embedding.

## IV. VIRTUAL NETWORK EMBEDDING MODEL AND PROBLEM FORMULATION

In this section, both substrate and virtual network models will be described then present VNE problem formulation with both cost and energy consumption as performance metrics.

### A. Network Model

- Substrate network (SN): This paper modeled the substrate network as an undirected weighted graph $Gs = (N_s, L_s, A_{s_n}, A_{s_l})$, where $N_s$, $L_s$ represent a set of substrate nodes and links respectively, and $A_{s_n}, A_{s_l}$ are the attributes of the nodes and links. We assumed that CPU represents the attributes of the nodes, while bandwidth represents the links attributes.

- Virtual network (VN): This represented as an undirected weighted graph $G_v = (N_v, L_v, R_{v_n}, R_{v_l})$, where $N_v, L_v$ represents a set of virtual nodes and links respectively, and $R_{v_n}, R_{v_l}$ are the requirements of the nodes and links denoted by the node required CPU and link required BW.

- Virtual network request (VNR): This paper represented a VN request as VNR $(G_v, T_a, T_d)$ where $T_a$ is the request arrival time, $T_d$ is the duration of VNR spent in SN and $G_v$ is the virtual network.

- Virtual network embedding (VNE): This is defined as the process of mapping the virtual nodes to the suitable substrate nodes, and mapping virtual links to the corresponding substrate paths in the SN. This mapping process should meet the node and link constraints as follows:

Node mapping: Mn :{i → J, u → K}

Link mapping: Ml :{ $L_v \to P_s$ }

Where, *i* and *u* are virtual nodes and *J* and *K* are substrate nodes that host virtual nodes. $P_s$ is part of the set that contains the substrate path in $G_s$. As you can see in (3), it illustrates both CPU and BW constraints, where the CPU provided by substrate node J should be greater than or equal to the CPU required by virtual node I, and the BW of path (J,K) should be greater than or equal to the BW required by the virtual link (i,u) in order to accommodate the VNR demands.

$$CPU(i) \leq CPU(J), BW(i,u) \leq BW(J,K), \forall i, u \in N_v \text{ and } \forall J, K \in N_s \quad (3)$$

*B. Energy Consumption Model*

As mentioned before, this paper focus on reducing the embedding cost along with energy consumption. In the presented model, the nodes in the SN will divide in two types. First, active nodes that participate in the current mapping process, as hosting nodes, or still active from the last one. For example, the active nodes in Fig. 1(b) are {A, B, E}. In contrast, idle nodes that still not participating in current mapping process, or already finished its job. Idle nodes in Fig. 1(b) are {C, D}. The energy consumption (E) for VNE process is the energy needed to power the nodes to on state plus the energy needed to host the virtual nodes.

$$E = \begin{cases} \sum_{i=1}^{u}(CPU(u).P_l(n_s)) + (P_b).(N) \\ \text{if node statues} = 0 \end{cases} \quad (4)$$

Where, N represents set of all substrate nodes that needed to power to ON state, and u represent set of all virtual nodes in VNR. $P_b$ represent server's baseline power and $p_l = p_u - p_b$. Where, $p_u$ represent energy consumed when the server reaches its highest utilization [10] [19].

*C. Performance Metrics*

*1) Energy consumption:* This is defined as the sum of all the power consumed on each substrate nodes during VNE process.

$$E(vnr) = \sum_{u \in N_v} E.T_d \quad (5)$$

And long term average energy consumption will be defined as following:

$$\lim_{T \to \infty} \frac{\sum_{i=1}^{N} E(vnr,t)}{T} \quad (6)$$

Where, T is process time for the whole VNR, and N is the number of accepted VNR.

*2) Embedding cost:* This is defined as the sum of all the CPU and BW requirements from the VNR during the VNE process.

$$C(vnr) = (\sum_{u \in N_v} CPU(u) + \sum_{l_{u,i} \in L_v} BW(l_{u,i})).T_d \quad (7)$$

From (5), and (7), we drive the conclusion of the objective functions in our paper that aim to minimize both "E" and "C".

$$\text{Min } (E + C) \quad (8)$$

*3) Embedding revenue:* This is the revenue of embedding ***vnri*** at time t, and is defined as the sum of all the substrate CPU and BW required by ***vnri***.

$$R(vnri,t) = (\sum CPU(u) \ u \in N_v + \sum BW(l_{u,i}))$$
$$l_{u,i} \in L_v \quad (9)$$

Where, $CP(u)$ is the required CPU by virtual node u, and $BW(L_{u,i})$ is the required bandwidth from virtual link $L_{u,i}$

V. PROPOSED ALGORITHM

*A. EAPSO_Single ISP*

*1) Particle position initialization:* The responsibility of this algorithm is finding near optimal solution for virtual network embedding in single ISP. It starts with particle's position initialization and ends with embedding solution. The particle's position represents candidate solutions for VNE problem. This algorithm takes substrate network as input along with virtual network. SN represented as the following: $G_s = (N_s, L_s)$. Where $N_s$ is a set of all substrate nodes in $G_s$, and $L_s$ is the set of all substrate links in $G_s$.

The algorithm starts with receiving virtual network request. Then constructing Breadth First Search over virtual network, starting with nodes with largest resources (CPU and BW), after constructing BFS tree, each level will be sort in descending order based on virtual nodes resources. At the end, we map virtual nodes starting from root node then from the next level. At SN, we start constructing candidate list for each virtual node: Candidate list constructed by creating BFS over SN starting with a node with highest resources. Then, sorting each BFS level in descending order based on total resources and finally collecting only substrate nodes that have resources higher or equal to requested virtual resources. By constructing BFS and choosing substrate nodes with enough BW to accommodate virtual nodes required BW, only nodes with enough BW will be considered, thus virtual link mapping will be considered at same time of creating substrate candidate list. After creating candidate list, initializing particle position vector stage will begin where each virtual node will choose from its candidate list, the host that could map to. Each node selected from the candidate list should update its remaining resource value.

*2) Energy:* Aware Virtual Network Embedding algorithm

This algorithm depends on checking the feasibility of each particle position. Particle considers feasible if there is at least one path in SN for every virtual link. If the particle is feasible, then it should be updated as the following:

$$v_i^d = p_1 v_i^d + p_2(pBest_i^d - x_i^d) + p_3(gBest - x_i^d) \quad (10)$$

$$x_i^d = x_i^d \times v_i^d \quad (11)$$

Otherwise, particle should be remapped in way that each virtual node will reselect nodes from its candidate substrate list as in subsection "Particle position initialization". After checking the feasibility, the fitness function of each particle should be calculated.

However, because of the nature of working in discrete workspace, PSO original operation should be modified to fit into presented Energy-Aware VNE algorithm. The modification will be as the following: The velocity will represent as vector generated randomly from each virtual node's candidate substrate list. Ex: $V = (v_1, v_2, v_3, v_i)$, where i represent the order of virtual nodes in VNR. Candidate substrate list of virtual node in dimension i contains neighbor nodes of substrate node in dimension i.

"X" represents the position vector.

Ex: $X = (x_1, x_2, x_3, x_i)$.

Where, i represent the order of virtual nodes in VNR.

- Subtraction (-): This operation depends on the fitness function of position $X_i$, pBest and gBest to apply exchanging rules. It works as by Calculate fitness function of $X_i$, pBest and gBest and then Seek for the conflict between two clauses. After that, change the values from the solution with worst fitness according to the position of the conflict values.

- Addition (+): $v_i p_i + v_j p_j$ indicate that substrate nodes will be kept from $v_i$ with probability of $p_i$, and kept from $v_j$ with probability of $p_j$. The selection technique we used called "Roulette Wheel Selection".

- Multiplication (*): This operation maps the virtual nodes that currently mapped to substrate nodes from $X_i$ to the corresponding substrate nodes in $V_{i+1}$. If the same dimension in $V_{i+1}$ and $X_i$ has the same node, the virtual nodes should reselect the substrate nodes from its candidate list as the following:

$$X_i * V_{i+1} = (A, B, A) * (A, D, C) = (N, D, C)$$

Where, N is new selected node from substrate candidate list.

Both position and velocity update will be as the following equations:

$$v_i^d = p_1 v_i^d \oplus p_2 \left(pBest_i^d \ominus x_i^d\right) \oplus p_3 \left(gBest^d \ominus x_i^d\right) \quad (12)$$

$$x_i^d = x_i^d * v_i^d \quad (13)$$

$p_1, p_2, p_3$ are random numbers generated in condition of: $p_1 + p_2 + p_3 = 1$

### B. EAPSO_Distributed Clouds

In this part of the algorithm VNR shall be partitioned to several sub-graphs to cope with problems consequent from lack of available resources or clients geographical constrains violation. VNR will be coarsened to smaller sub-graphs then assigning phase will start. This algorithm composed of three main methods: Coarsening graph, Un-coarsening graph, and Construct graph (Algorithm 1).

*1) Coarsening VNR phase*: In this phase, the original graph will be coarsened to several yet smaller sub-graphs. Each of these sub-graphs composed of several vertices grouped in single node. The weight of coarsened node equal to the sum of vertices contained. The proposed algorithm follows Heavy Clique Matching (HCM) to obtain coarsen graphs [16], [20].

*2) Uncoarsening phase:* Before sending the coarsen node to suitable DC, each node should be projected to its initial graph. In this case the coarsen node assigned to DC in form of sub-graph. Uncoarsening function responsible of finding all virtual nodes that collapsed inside the coarsen node $C_n$. The obtained nodes will pass to Construct_graph function.

*3) Construct graph phase*: In this phase, each coarsen node will be reconstructed as graph before it sent to the suitable DC.

As seeing in Algorithm2, using same approach as the one used in single ISP, it starts with constructed BFS tree from the coarsen graph. In the created BFS tree, each coarsen nodes in each level will be sorted in descending order based on their requested resources. Starting from root node, each coarsens node will reconstructed by creating sub-graph from uncoarsen nodes and links connecting them. Each sub-graph will be assigned to SN that copes with its requirements. If assign function ends, each SN will start mapping its assigned VN as single ISP scenario, otherwise Algorithm 2 will terminate and assign_flag return false.

---

**Algorithm 1**: *Energy-Aware Virtual Network Embedding Algorithm*

---

*Input:* virtual network $G_v = (N_v, L_v)$.

Substrate network $G_s = (N_s, L_s)$.

*Output:* Embedding solution.

**Begin**

1: Initializing particles population.
2: Initializing pBest and gBest for each particle.
3: While stopping criteria in not satisfied.
4: For each particle i do {
5: Get fitness function $f(x_i)$ for particle I.
6: If($particle\ position.feasable()$) then {
7: Update particle position and velocity according to Equations (12), (13).
8: Else
9: Re- initializing its position, and recreate velocity vector randomly.
10: } End if
11: } End for
12: Get $pBest_i$ and $gBest$
13: If $f(x_i) < f(pBest_i)$ then
14: Set $x_i$ as new personal best position for particle i.
15: If $f(pBest_i) < f(gBest)$ then
16: set $pBest_i$ As new global best position.
17: End if
18: End while
19: Embedding solution is final $gBest$.

**End**

**Algorithm 2**: *EAPSO_distributed clouds*

Input: virtual network to embed $G_v = (N_v, L_v)$.

Substrate networks $G_s = (N_s, L_s)$.
$N_s$ is substrate network to assigned to.

$L_s$: All substrate links between substrate networks.

Min_resource: minimum available resource in SN.

Output: assign_flag: VNE success flag.

**Begin**

1: $G_c$=coarsen ($G_v$,min_resource).
2: Build BFS tree of $G_c$ starting with coarsen node with highest resources as root.
3: Sort nodes in each level in descending order based on their required resource.
4: For each node $N_c$ in the sorted list construct sub_graph after obtaining it's uncoarsen nodes.
5: Un_coarsening()
6: Sub_G=Construct_graph ($N_c$).
7: If (assign (sub_G,$G_s$)).
8: Then.
9: assign_flag=true.
10: Return.
11: Else.
12: Assign _flag=false.
13: Return
14: End if
15: End for
**End**

## VI. EVALUATION

In order to support the presented assumption and evaluate the efficiency of the proposed algorithm, extensive experiments conducted to evaluate proposed algorithm against the following algorithms: HCM [15], RW-MaxMatch [16], BFSN-HEM [17], AdvSubgraph-MM [13], BFSN [17], AdvSubgraph-MM-EE [13], and AdvSubgraphMM-EE-Link [13].

### A. Environment Setting

Using Waxman generator, substrate network topology is generated with 100 nodes and 500 links. Bandwidth of the substrate links are uniformly distributed between 50 and 150 with average 100. Each substrate node is randomly assigned one of the following server configurations: HP ProLiant ML110 G4 (Intel Xeon 3040, 2 cores X 1860 MHz, 4 GB), or HP ProLiant ML110 G5 (Intel Xeon 3075, 2 cores X 2660 MHz, 4 GB). In addition, 1000 Virtual network topologies are generated using Waxman generator with average connectivity 50%. The number of virtual nodes in each VN is variant from 2 to 20. Each virtual node is randomly assigned one of the following CPU: 2500 MIPS, 2000 MIPS, 1000 MIPS, and 500 MIPS, which are correspond to the CPU of Amazon EC2 instance types. Bandwidths of the virtual links are real numbers uniformly distributed between 1 and 50. VN's arrival times are generated randomly with arrival rate 10 VNs per 100 time units. The lifetimes of the VNRs are generated randomly between 300 and 700 time units with average 500 time units. Maximum allowed hop is set to 2 and maximum backtrack is set to 3*n, where n is the number of VN nodes. Generated SN and VNs topologies are stored in brite format and used as inputs for all mapping algorithms.

### B. Experiment Results

In case of energy consumption, it represents energy consumption comparison according to achieved resources starting from t=5000 until t=10000. Taking look in Fig. 2, it shows that energy at time T=5000 equals to 1 kilowatt and start to fall until T=9000 where its reach 0.8 kilowatt. By observing energy at T=7000 and T=9000 the algorithm presents rise of consumption at T=9000 even though acceptance ratio in both time is equal to 29% (Fig. 3). The reason behind this variant in energy consumption goes to the increase in number of achieved resources at T=9000 with 13.23% (Fig. 4). On the other hand, by comparing proposed algorithm with previous VNE algorithms, it can be noticed that EAVNE_PSO scores lowest consumption among all.

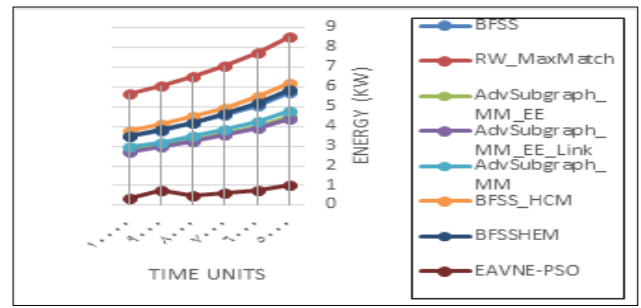

Fig. 2. Energy consumption.

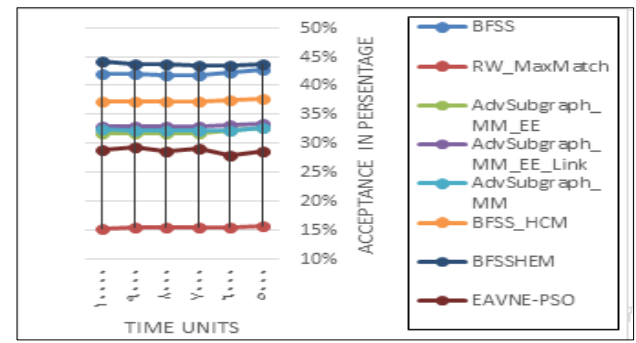

Fig. 3. Acceptance comparison.

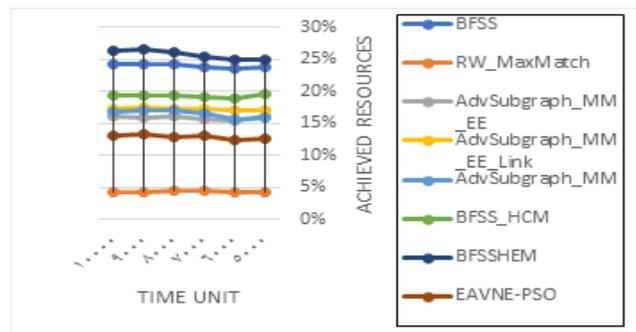

Fig. 4. Achieved resource comparison.

Fig. 5 shows the regression of long term average revenue as time increase, this goes to the fact that while time pass, more VNR will finished and depart from SN, thus decreasing overall revenue at that point. In comparison to the previous VNE algorithms, EAVNE_PSO outperforms RW-MaxMatch with 1000 "in money unit" while maintaining below the remaining algorithms.

From Fig. 6, the proposed algorithm at T=5000 score highest value of R/C over 96% which is the same time of highest registered revenue. While by comparing value at T=6000 and T=7000 it is clear that even T=6000 score higher revenue than T=7000, it encounters drop of R/C value. This drop can be explained that cost of mapping VNRs at T=6000 is higher than one in T=7000. From the previous, it noticed that highest R/C means smaller gap between value of revenue and cost and vice versa. Fig. 6 presents that proposed algorithm R/C values did not exceed any algorithms that involved in the experiment. In order to improve revenue over cost, proposed algorithm allows more than one virtual nodes to map on same substrate node, thus decrease cost of mapping these nodes since cost of mapping link between them is 0.

EAVNE_PSO has significant increase comparing to RW-MaxMatch with 15% acceptance ratio while still not outperforms the remaining algorithms witch reach to 44% as in BFSSHEM. The reason behind the rejection of VNRs that reach to over 85% as in Fig. 7 goes to the lack of available resources either CPU or BW as showing in Fig. 8 and 9. As showing in Fig. 3, at time 9000 the availability of CPU decreases due to the fact that in that time the acceptance ratio increased to 28%. From this point, the proposed algorithm highlights the fact that while experiment time increase, the acceptance ratio will rise due to the fact that by time passing more VNRs will depart the SN and therefore the resources availability will increase.

Even that it looks like proposed algorithm has low acceptance ratio among other VNE algorithms (except RW-MaxMatch), it must be emphasized on fact of the heuristic nature of the proposed algorithm in finding near optimal solution instead of exact one. It gives the proposed algorithm the advantage of working in larger networks (more than 200 nodes) and find near optimal solution in reasonable embedding time, in contrast of exact solution algorithm which take exponential time to find solution. Therefore, proposed algorithm will be even more efficient in larger networks.

From Fig. 10, the VNE time for the proposed algorithm varies between 4300 and 4500 milliseconds. This range falls between times of AdvSubgraph-MM, AdvSubgraph-MM-EE and AdvSubgraph-MM-EE-Link which score highest VNE time reach to 50000 milliseconds in AdvSubgraph-MM-EE-Link algorithm and times of the remaining algorithms that start from 300 milliseconds in BFSS and drops from there.

Through further observation in EAVNE_PSO VNE time, it perceives that times have almost steady rate with slight variation. This is due to the nature of EAVNE-PSO algorithm in finding near optimal solution. The proposed algorithm goes through the same steps and iterations for every VNR. It starts from creating fixed number of candidate solutions and run them through fixed number of iterations.

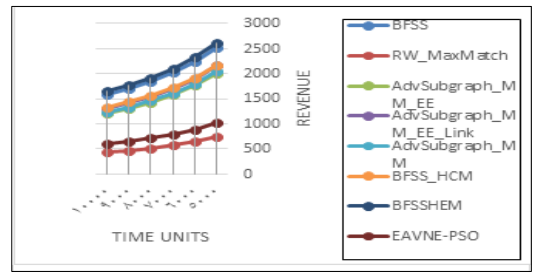

Fig. 5. Revenue comparison.

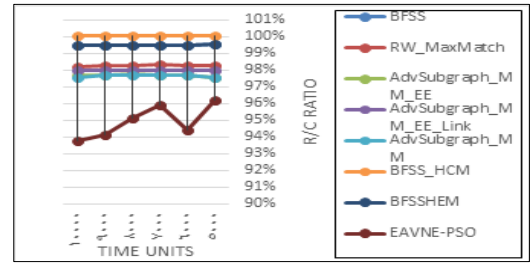

Fig. 6. R/C comparison.

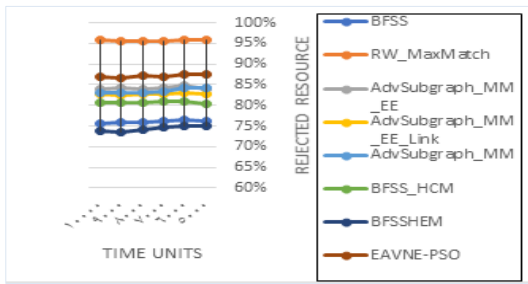

Fig. 7. Rejected resource comparison.

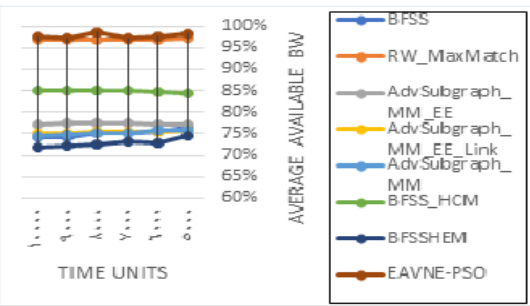

Fig. 8. Average available CPU comparison.

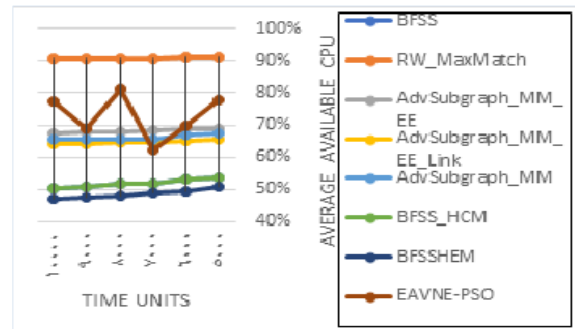

Fig. 9. Average available BW comparison.

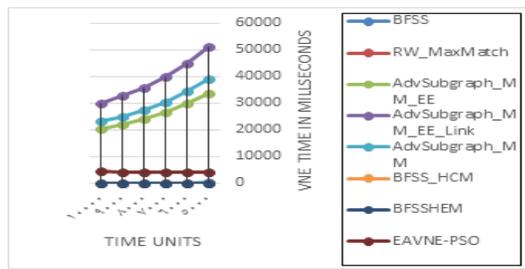

Fig. 10. VNE time comparison.

## VII. CONCLUSION

This paper tackled the importance of cloud computing and network virtualization to fulfill the growth demands for resources provisioning. One of the most important issues in resources provisioning is how to map virtual resources to corresponding physical one. This issue defined as VNE problem. VNE considered as NP-hard problem and it gain a lot of attention last few years. This paper proposed Energy Aware Virtual Network Embedding in Distributed Clouds. The proposed algorithm aims to find near optimal solution based on modified Particle Swarm Optimization algorithm. For the distribution purpose, VNR will be partitioned in to sub-graphs. After partitioning phase is over each sub-graph will be sent to the suitable DC where modified PSO is initiated.

The proposed modified PSO reduce both embedding cost and energy consumption by finding embedding solution that contains least idle nodes that needs to be powered on. In addition, by virtual node consolidation along with setting maximum allowed hop to 2 to prevent the situation of placing two neighbor nodes far away from each other, thus reducing embedding cost.

After conducting extensive experiments to evaluate the efficiency of the proposed algorithms, it shows that EAVNE-PSO algorithm outperform RW-MaxMatch algorithm in energy consumption, revenue, acceptance ratio, achieved and rejected resources. After extensive observation, the proposed algorithm achieved acceptance ratio that represent 70% of highest acceptance ratio among the tested previous works "BFSSHEM", while energy consumption is 9% less. In addition, the number of accepted VNRs along with the size of accepted VN has a significant impact on revenue gained and energy consumption path splitting technique in link mapping is planned to be adopted in the future work. This technique will decrease rejected BW by partitioning a single virtual link BW to be mapped through different substrate paths. Furthermore, virtual nodes migration will add further improvements in energy and cost preservation. By moving virtual nodes from under- or over-utilized hosts, it can increase the chances of switching underutilized hosts to idle mode and consolidating virtual nodes in the same host.